\documentclass[showpacs,aps,prb,floatfix,twocolumn,superscriptaddress]{
revtex4-1}
\usepackage[pdftex]{graphicx}
\usepackage{dcolumn}
\usepackage{bm}
\usepackage{amsmath}
\usepackage{array}
\usepackage{color}
\usepackage{float}
\usepackage{subfigure}
\usepackage{dsfont}
\usepackage{txfonts}
\usepackage{wasysym}
\usepackage{multirow}
\usepackage{sidecap}
\usepackage{xcolor,cancel}
\usepackage[normalem]{ulem}

\newcommand{\D}{\Delta}
\newcommand{\+}{\dagger}

\newcommand{\e}{\varepsilon}

\newcommand{\leads}{\mathrm{leads}}
\newcommand{\QD}{\mathrm{dot}}
\newcommand{\QDleads}{\mathrm{dot-leads}}
\newcommand{\chain}{\mathrm{chain}}
\newcommand{\QDchain}{\mathrm{dot-chain}}

\newcommand{\meV}{~\text{meV}}
\newcommand{\mueV}{~\mu\text{\rm eV}}

\newcommand{\veck}{\mathbf{k}}

\begin{document}



\title{Subtle leakage of a Majorana mode into a quantum dot}

\author{E. Vernek}
\affiliation{Instituto de F\'isica, Universidade Federal de Uberl\^andia, 
Uberl\^andia, Minas Gerais 38400-902, Brazil.}
\affiliation{Instituto de F\'{i}sica de S\~ao Carlos, Universidade de S\~ao 
Paulo, S\~ao Carlos, S\~ao Paulo 13560-970, Brazil}

\author{P.~H.~Penteado}
\affiliation{Instituto de F\'{i}sica de S\~ao Carlos, Universidade de S\~ao 
Paulo, S\~ao Carlos, S\~ao Paulo 13560-970, Brazil}

\author{A.~C.~Seridonio}
\affiliation{Departamento de F\'isica e Qu\'imica, Universidade Estadual 
Paulista, Ilha Solteira, S\~ao Paulo 15385-000, Brazil}

\author{J.~C.~Egues}
\affiliation{Instituto de F\'{i}sica de S\~ao Carlos, Universidade de S\~ao 
Paulo, S\~ao Carlos, S\~ao Paulo 13560-970, Brazil}

\date{\today}

\begin{abstract}
We investigate quantum transport through a quantum dot connected to source 
and drain leads and side-coupled to a topological superconducting nanowire 
(Kitaev chain) sustaining Majorana end modes. Using a recursive Green's 
function approach, we determine the local density of states (LDOS)
of the system and find that the end Majorana mode of the wire leaks into 
the dot thus emerging as a unique dot level {\it pinned} to the Fermi 
energy $\e_F$ of the leads. Surprisingly, this resonance pinning, 
resembling in this sense a ``Kondo resonance'', occurs even when the 
gate-controlled dot level $\e_\text{dot}(V_g)$ is far above or far below 
$\e_F$.  The calculated conductance $G$ of the dot exhibits an unambiguous 
signature for the Majorana end mode of the wire: in essence, an 
off-resonance dot [$\e_\text{dot}(V_g)\neq \e_F$], which should have $G=0$, 
shows instead a conductance $e^2/2h$ over a wide range of $V_g$, due to 
this pinned dot mode. Interestingly, this pinning effect only occurs when 
the dot level is coupled to a Majorana mode; ordinary fermionic modes  
(e.g., disorder) in the wire simply split and broaden (if a continuum) the 
dot level. We discuss experimental scenarios to probe Majorana modes in 
wires via these leaked/pinned dot modes.
\end{abstract}

\date{\today}
\maketitle

\section{Introduction} 
Zero-bias anomalies in transport properties 
are one of the most intriguing features of the low-temperature physics in  
nanostructures.  The canonical example is the zero-bias peak in the  
conductance of interacting quantum dots coupled to metallic contacts, which 
is a clear manifestation of the Kondo 
effect~\cite{Nature.391.156,Science.S.1998.540} arising from the dynamical 
screening of the unpaired electron spin in the quantum dot by the itinerant 
electrons of the leads. Another example is the Andreev bound state arising 
from electron and hole scattering at a normal-superconductor 
interface.\cite{PhysRevLett.103.077003}
\begin{figure}[!htb]
\centerline{\resizebox{3.3in}{!}{
\includegraphics{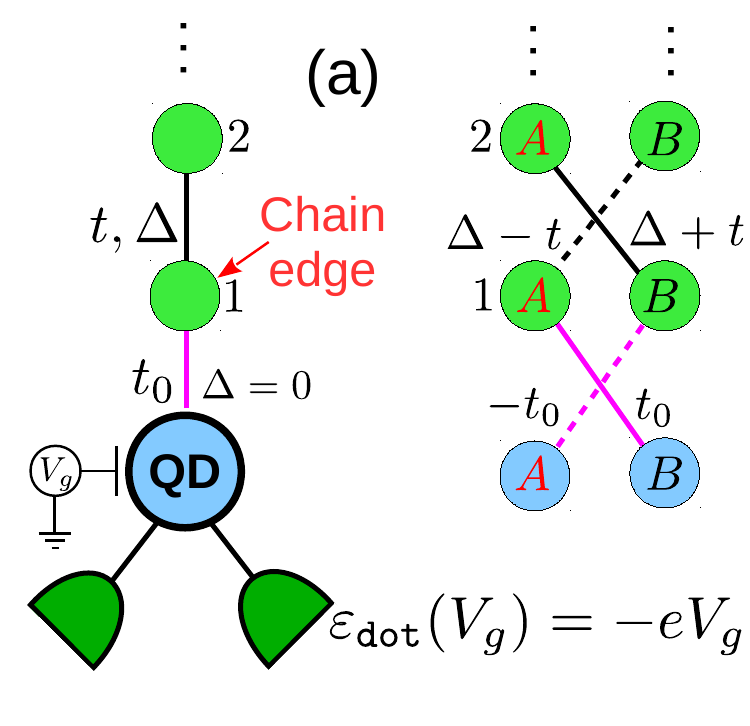}\hskip0.25cm\includegraphics{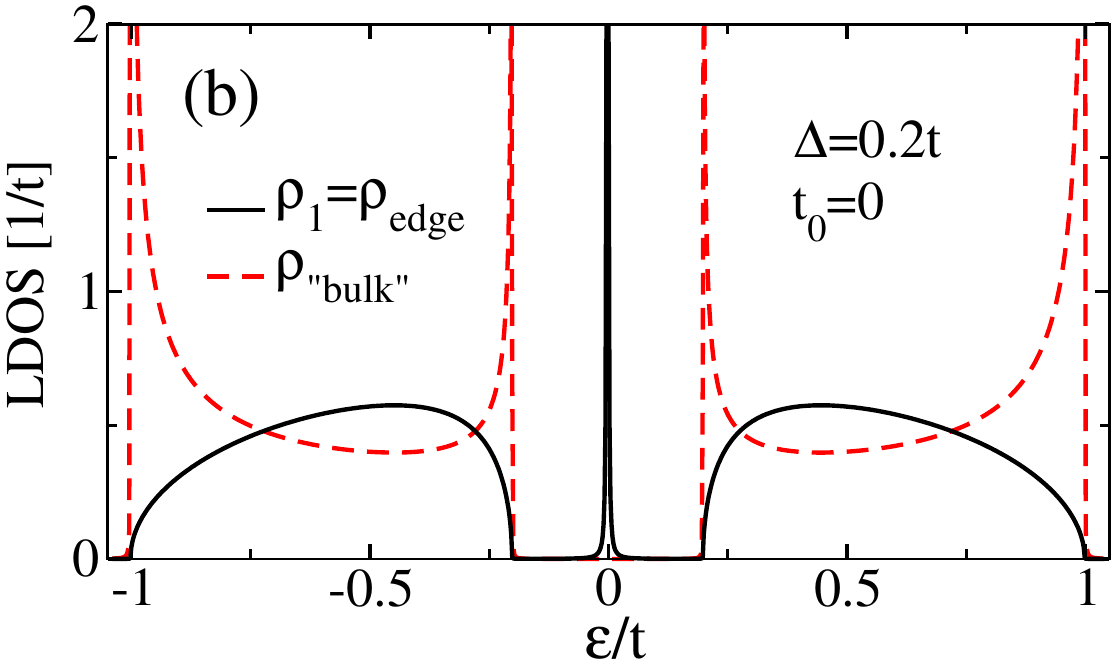}}}
\vskip-0.01in
\centerline{\resizebox{3.4in}{!}{\hskip0.15in\includegraphics{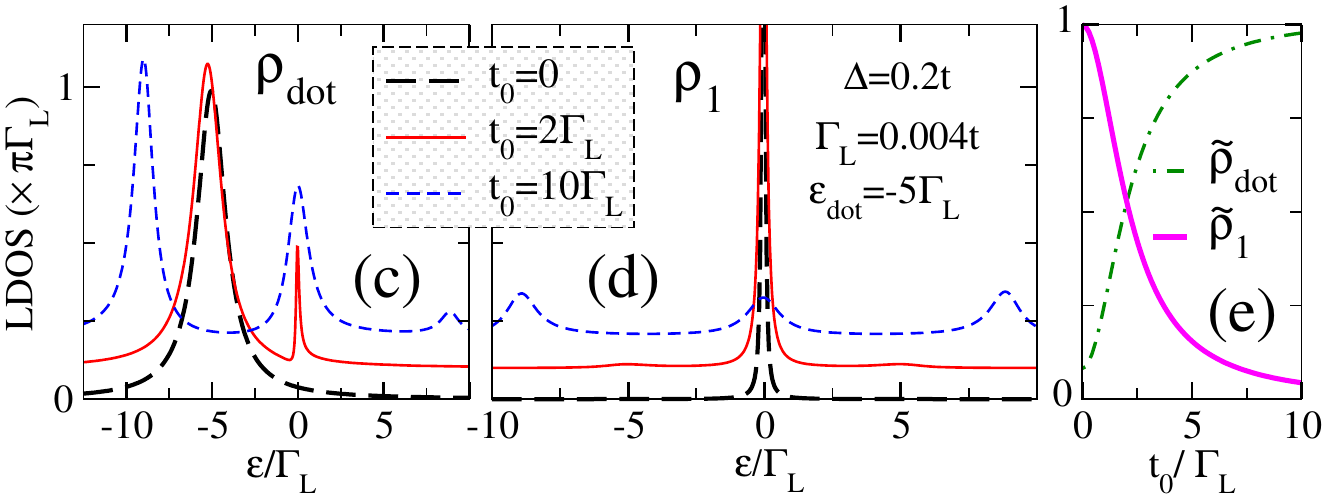}}}
\vskip-0.05in
\centerline{\resizebox{3.45in}{!}{\hskip0.15in
\includegraphics{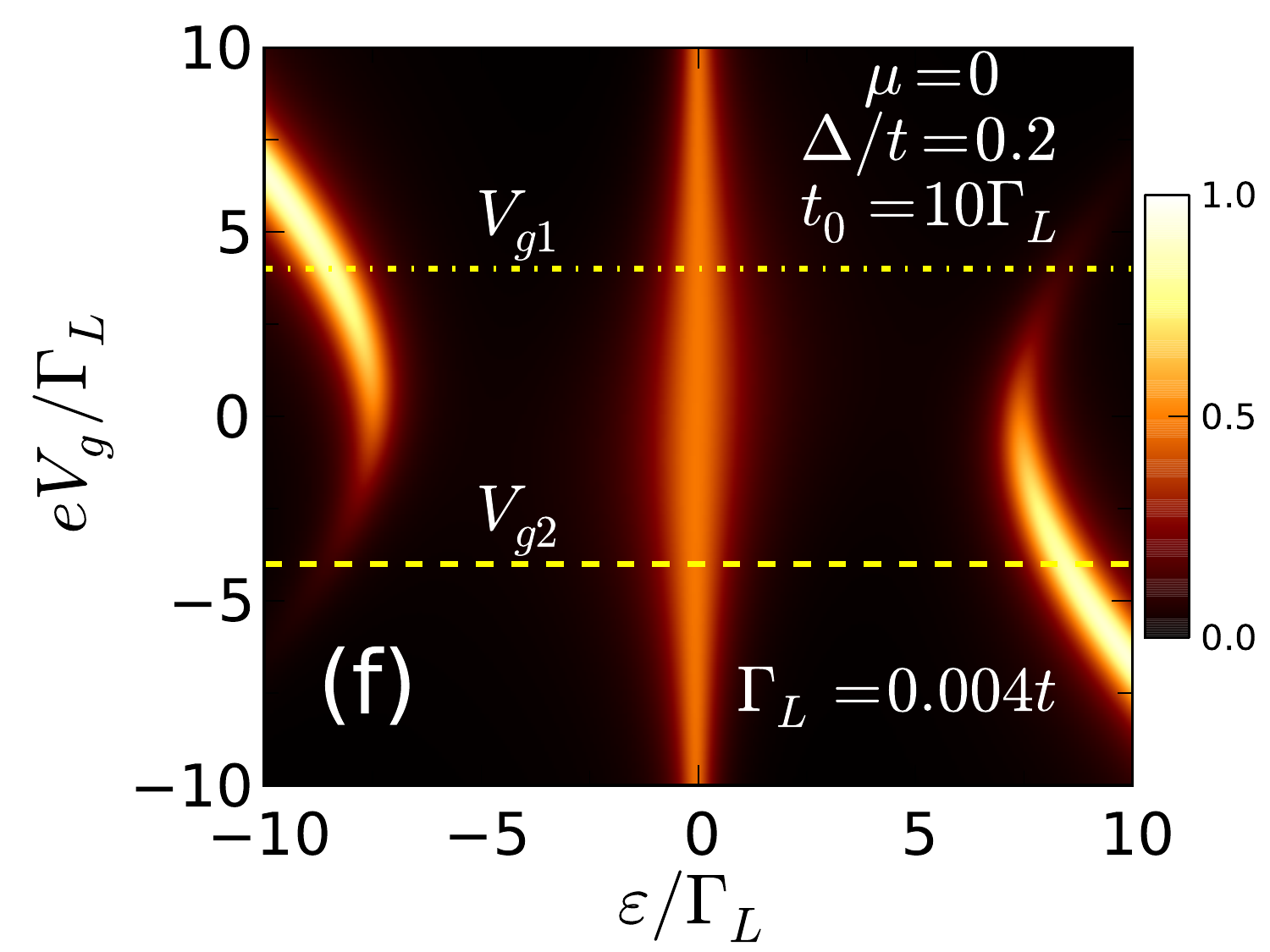}\includegraphics{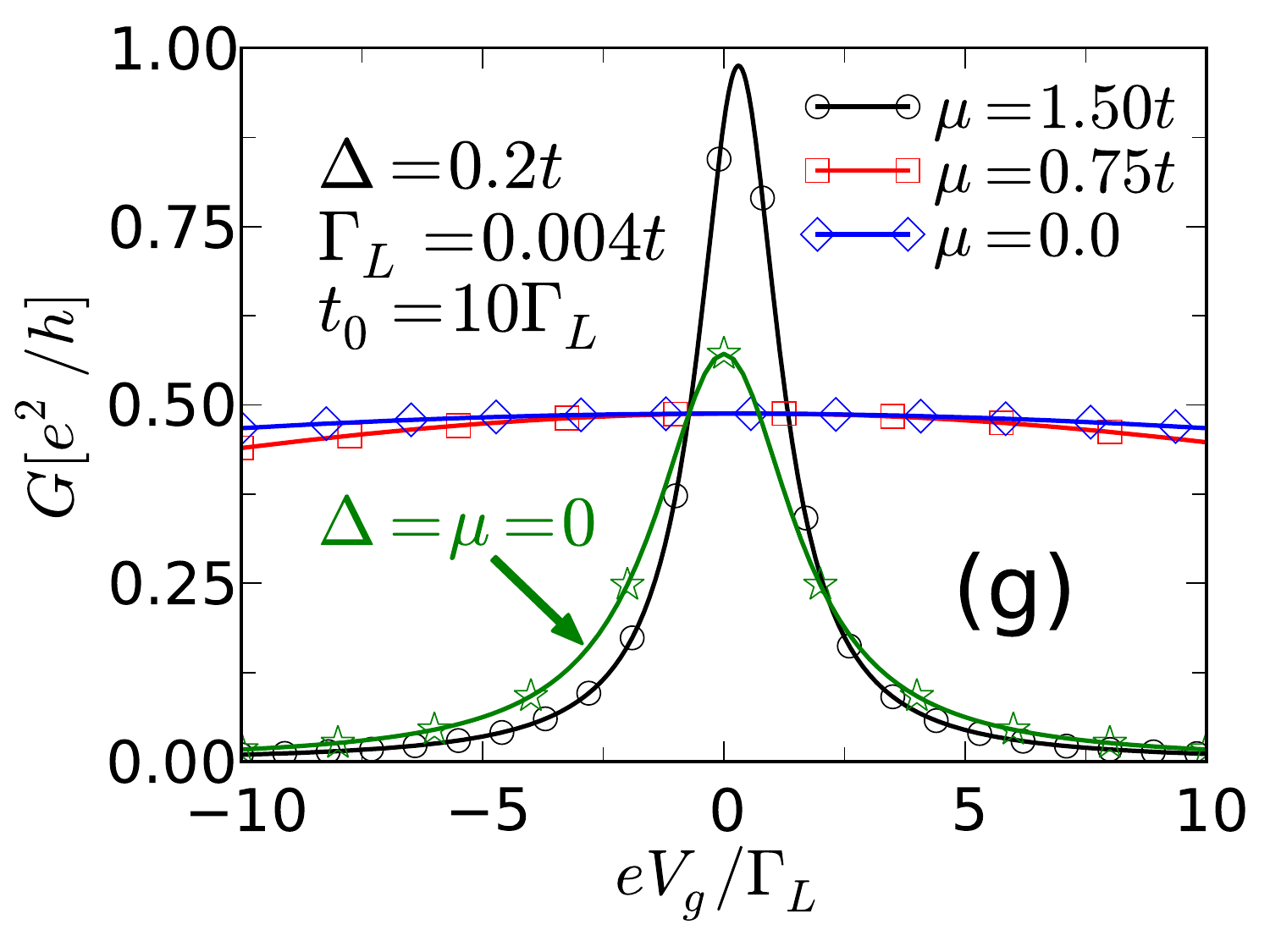}}}
\centerline{\resizebox{3.3in}{!}{
\hskip0.8cm\includegraphics{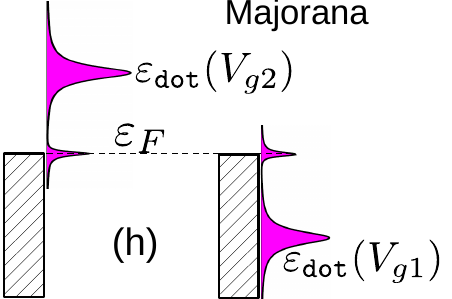}\hskip0.1in\includegraphics{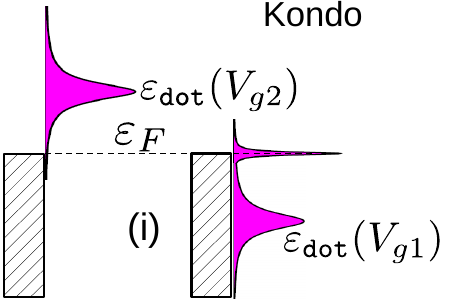}}}	
\caption{\label{fig1} (Color online) (a) Illustration of (left) a 
quantum dot (QD) side-coupled to a Kitaev wire and to two metallic leads 
and (right) the Majorana representation of the dot and the Kitaev chain. 
(b) ``Bulk" [dashed (red) line] and edge [solid (black) line] chain LDOS 
for $t=10\meV$, $\mu=0$, $\Delta=2\meV$, $\Gamma_L=40\mueV$ and $t_0=0$. 
LDOS of the dot $\rho_{\rm dot}$ (c) and of the first site of the Kitaev 
chain $\rho_1$ (d) for the same set of parameters as in (b) and various 
values of $t_0$. For clarity, the curves in (c) and (d) are offset 
along the $y$-axis. (e) $\tilde\rho_{\rm dot}=\rho_{\rm dot}(0) /\rho^{\rm 
max}_{\rm dot}$ and $\tilde\rho_{1}=\rho_{\rm dot}(0)/\rho^{\rm max}_{1}$ 
at $\e=0$ as functions of $t_0$, in which $\rho^{\rm max}_{\rm dot,1}={\rm 
max}[\rho_{\rm dot,1}(\e=0,t_0)]$. (f) Color map of the LDOS of the dot vs  
$\e$ and $eV_g$. (g) Conductance $G$ vs $eV_g$ for the same set of  
parameters as in (b) for various values of $\mu$. For comparison we show 
the case  $\Delta=\mu=0$ [stars (green)]. In (h) and (i) we sketch the LDOS 
of the dot for the Majorana and Kondo cases, respectively.}
\end{figure}

Recently, a new type of zero-bias anomaly has emerged in connection with 
the appearance of Majorana bound states in Zeeman split nanowires with 
spin-orbit interaction in close proximity to an s-wave 
superconductor.\cite{PhysRevLett.105.077001,PhysRevLett.105.177002} It is  
theoretically well established that these ``topological'' superconducting 
wires sustain chargeless zero-energy end states with peculiar features such 
as braiding statistics, possibly relevant for topological quantum 
computation.\cite{Ann.Phys.303.2,RevModPhys.80.1083} Experimentally, 
however, there is still controversy as to what the observed zero-bias peak 
really means: Kondo effect, Andreev bound states and disorder effects are 
some of the 
possibilities.\cite{Law,Pikulin,Science.336.1003,NanoLetters.12.6414-6419, 
Nat.Phys..8.887,PhysRevLett.109.186802, 
PhysRevLett.110.126406,PhysRevB.87.241401} Franz summarizes and discusses 
these issues in Ref.~\onlinecite{franz}.

Here we propose a direct way to probe the Majorana end mode arising in a 
topological superconducting nanowire by measuring the two-terminal 
conductance $G$ through a dot side-coupled to the wire, 
Figs.~\ref{fig1}(a) and \ref{fig1}(b). Using an exact recursive 
Green's function approach, we calculate the local density of states (LDOS) 
of the dot and wire, and show that the Majorana  end mode of the 
wire leaks into the dot~\cite{footnote2} thus giving rise to a Majorana 
resonance in the dot, Figs.~\ref{fig1}(c) and \ref{fig1}(d). Surprisingly, 
we find that this dot Majorana mode is pinned to the Fermi level $\e_F$ of 
the leads even when the gate controlled dot level $\e_{\rm dot}(V_g)$ is 
far off resonance $\e_\text{dot}(V_g) \neq  \e_F$. 

Based on the results above, we suggest three experimental ways of probing
the Majorana end mode in the wire via the leaked/pinned Majorana mode in 
the  dot: (i)  with the dot kept off resonance [$\e_\text{dot}(V_g) \neq 
\e_F$] one can measure $G$ vs $t_0$, the wire-dot coupling $t_0$ can be 
controlled by an external gate, to see the emergence of the $e^2/2h$ 
peak in $G$ as the Majorana end mode ``leaks'' into the dot, Fig.~\ref{fig1}(e) 
(cf. $\rho_{\rm dot}$ and $\rho_1$, see also Fig.~\ref{fig2}); (ii) 
Alternatively, one can measure $G$ vs $V_g$ over a range in which $\e_{\rm 
dot}(V_g)$ runs from far below to far above the Fermi-level of the leads 
where we find $G$ to be essentially a {{\it plateau}} at $e^2/2h$, 
Figs.~\ref{fig1}(f) and \ref{fig1}(g); 
(iii) Yet another possibility is to 
drive the wire through a non-topological/topological phase transition, 
e.g., electrically via the spin-orbit coupling, temperature or 
the chemical potential $\mu$ of the wire (Fig.~\ref{fig3}), while 
measuring the conductance of the dot; the presence/absence of the Majorana 
end mode in the wire would alter drastically the conductance of the dot,  
see circles (black) and stars (green) in Fig.~\ref{fig1}(g).

The above pinning of the dot Majorana resonance at $\e_F$ is  similar to 
that of Kondo.\cite{Hewson-Kondo} However, the Kondo resonance only occurs 
for $\e_{\rm dot}(V_g)$  {\it below}  $\e_F$ [cf. Figs.~\ref{fig1}(h) and 
\ref{fig1}(i)] and yields a conductance peak at $e^2/h$ (per spin) instead. Even 
though there is no Kondo effect in our system (spinless dot), 
we conjecture that this symmetry of the dot-Majorana resonance with 
respect to $\e_{\rm dot}(V_g)$ above and below $\e_F$ could be used to 
distinguish Majorana-related peaks from those arising from the usual Kondo 
effect whenever this effect is relevant.\cite{note-1} Moreover, this 
Majorana resonance in the dot follows quite simply by viewing the dot as an 
additional site (though with no pairing gap) of the Kitaev 
chain.\cite{Phys.-Usp..44.131,NJP045020} We emphasize that this unique 
pinning occurs only when  the dot is coupled to a Majorana mode -- a 
half-fermion state. When the dot is coupled to usual fermionic modes 
(bound, e.g., due to disorder, or not) in the wire, its energy level will 
simply split and broaden as we discuss later on. A spin full version of our 
model with a Hubbard $U$ interaction in the dot yields similar 
results.\cite{mapping-U}

{The paper is organized as follows. In Secs.~\ref{Hamiltonian} and 
\ref{Green_Funcions} we present the Hamiltonian that describes our system 
and introduce the Majorana Greens functions that we use to calculate
the relevant physical quantities, respectively. In Sec.~\ref{Results} we
present our numerical results and discussions. Finally,  we summarize 
our main findings in Sec.~\ref{Remarks}.}

\section{Model Hamiltonian}\label{Hamiltonian}
We consider a single-level spinless quantum dot  coupled to two  metallic 
leads and to a Kitaev chain,\cite{mapping-U}  
Fig.~\ref{fig1}(a). {To realize a single-level dot (`spinless' dot regime), 
we 
consider a dot with gate controlled Zeeman-split levels 
$\varepsilon_{dot}^\downarrow(V_g) = -eV_g$ ($e>0$)  and  $\varepsilon_{dot}^\uparrow(V_g) = \varepsilon_{dot}^\downarrow(V_g) + 
V_Z$, with $V_Z$ the Zeeman energy. By varying $V_g$ such that $|eV_g| < V_Z/2$ we can maintain the dot 
either empty [i.e., both spin-split levels above the Fermi level 
$\varepsilon_F$ (taken as zero) of the leads]  or singly occupied [i.e., 
only one spin-split dot level, e.g., $\varepsilon_{dot}^\downarrow(V_g)$ 
below $\varepsilon_F$]. This is the relevant spinless regime in our setup.\cite{note-2} 
Typically [e.g.,  Fig.~\ref{fig1}(g)], we vary {$|eV_g| < 10\Gamma_L=0.4 \meV$}, 
assuming a realistic Zeeman energy to attain topological 
superconductivity, i.e., {$V_Z \simeq 0.8 \meV$} [see Rainis et al. in Ref. 
\onlinecite{PhysRevB.87.024515}]. This picture also holds true in the 
presence of  a Hubbard $U$ term in the dot cite{mapping-U}). 
In this spinless regime, our} Hamiltonian is $H=H_{\chain}+H_{\QD}+ H_{\QDchain} 
+ H_{\leads}+ H_{\QDleads}$, with $H_{\chain}$ describing the chain 
\begin{equation}\label{H_wire}
H_{\mathrm{chain}}=-\mu\sum_{j=1}^{N}c_j^\+c_j{-}\frac{1}{2}\sum_{
j=1}^{N-1 } \left [tc_{j} ^\+c_ {j+1} +\D e^{ i\phi} c_ {j} c_{j+1} 
+H.c.\right],
\end{equation}
$N$ is the number of chain sites, $c^\dagger_j$ ($c_j$) creates 
(annihilates) a spinless electron in the $j-$th site and $\phi$ is an 
arbitrary phase. The parameters $t$ and $\Delta$ denote the inter-site 
hopping and the superconductor pairing amplitude of the Kitaev model, 
respectively; its chemical potential is $\mu$.

The single-level dot Hamiltonian $H_{\QD}$ is
\begin{eqnarray}
H_{\QD}=\left(\e_\QD-\e_F\right) c^\dagger_0 c_0, 
\end{eqnarray}
$c^\dagger_0$ ($c_0$) creates (annihilates) a spinless electron in 
the dot {with energy $\e_\QD=-eV_g$}  
and $H_{\leads}$ denotes the free electron source ({\it S}) and drain ({\it 
D}) leads
\begin{eqnarray}\label{H_lead}
 H_{\leads}=\sum_{\veck, \ell =S,D}(\e_{\ell,\veck}-\e_F) c^\dagger_{\ell 
,\veck}c_{\ell,\veck},
\end{eqnarray}
where $c^\dagger_{\ell,\veck}$ ($c_{\ell,\veck}$) creates (annihilates) a  
spinless electron with wavevector $\veck$ in the leads,  whose  Fermi level 
is $\e_F$. The coupling between the QD and the first site of the chain and 
between the QD and the leads are, respectively, 
\begin{eqnarray}\label{H_dot-chain}
 H_{\QDchain}=t_0\left(c^\dagger_0 c_{1}+c_{1}^\dagger c_0\right)
\end{eqnarray}
and 
\begin{eqnarray}\label{H_dot-leads}
 H_{\QDleads}=\sum_{\veck,\ell =S,D}\left(V_{\ell, \veck}c^\dagger_0 
c_{\ell,\veck}+H.c.\right).
\end{eqnarray}
{The quantity} $V_{\ell,\veck}$  is the tunneling between the QD and the 
source and drain leads and $t_0$ is the hopping amplitude between the QD 
and the Kitaev chain.

\section{Recursive Green's function and LDOS} \label{Green_Funcions}
Our model and approach are similar to those of 
Ref.~\onlinecite{PhysRevB.84.201308} and go beyond low-energy effective 
Hamiltonians.\cite{Martin} Let us introduce the Majorana fermions 
$\gamma_{\alpha j}$, $\alpha = A,B$, via 
$c_{j}={e^{-i\phi/2}}(\gamma_{Bj}+ i\gamma_{Aj})/2$ and $c^\+_{j} 
={e^{i\phi/2}}(\gamma_{Bj}-i\gamma_{Aj})/2$, $j=0\cdots N$ ($j=0$ is the 
dot).\cite{Phys.-Usp..44.131,0034-4885-75-7-076501} The $\gamma_{\alpha 
j}$'s  obey $[\gamma_{\alpha j}, \gamma_{\alpha^\prime j^\prime}]_+ 
=2\delta_{\alpha\alpha^\prime} \delta_{jj^\prime}$ and $\gamma_{\alpha 
j}^\dagger = \gamma_{\alpha j}$. We now define the Majorana retarded 
Green's function
\begin{eqnarray}\label{GF-0}
 {M}_{\alpha i,\beta j}(\e)=-i\int_{-\infty}^\infty\Theta(\tau)\langle 
[\gamma_{\alpha i}(\tau),\gamma_{\beta j}(0)]_+\rangle e^{i\e(\tau)}d\tau,
\end{eqnarray}
where $\langle \cdots \rangle$ 
denotes either a thermodynamic average or a ground state 
expectation value at zero temperature; $\Theta(x)$ is the Heaviside 
function and $\e\rightarrow \e+i\eta$, with $\eta\rightarrow 0^+$. 
We can express the electron Green's function as
\begin{eqnarray}\label{EGF}
 {G}_{ij} (\e)=\frac{1}{4} \left[{M}_{Ai,Aj} + {M}_{Bi,Bj}(\e) 
+i\left({ M}_{Ai,Bj}-{M}_{Bi,Aj}\right ) \right]
\end{eqnarray}
and determine the electronic LDOS
$\rho_j(\e)=(-1/\pi)\text{Im }  G_{jj}(\e)$, 
\begin{equation}\label{LDOS}
  {\rho}_{j}(\e)=\frac{1}{4}\left[{\cal A}_{j}(\e)+{\cal 
  B}_{j}(\e)-\frac{1}{\pi}{\tt
  Re}\left[{M}_{A{j},B{j}}(\e)-{M}_{B{j},A{j}}(\e)\right]\right].
\end{equation}
In~(\ref{LDOS}) we have introduced the Majorana LDOS ${\cal 
A}_{j}(\e)=(-1/\pi)\text{Im } {M}_{A{j},A{j}}(\e)$ and ${\cal 
B}_{j}(\e)=(-1/\pi)\text{Im } {M}_{B{j},B{j}}(\e)$.

Using the equation of motion for the Green's functions, we obtain a 
set of coupled matrix equations, e.g., for $j=0$ (dot)
\begin{eqnarray}\label{G11}
 \mathbf{M}_{00}(\e)&=&\bar{\bf m}_{00}(\e)+\bar{\bf m}_{00}(\e){\bf
W}^\dagger_0{\bf M}_{10}(\e),
\end{eqnarray}
 where ${\bf M}_{ij}(\e)$ is [see Eq.~(\ref{GF-0})]
\begin{eqnarray}\label{G_ij_Matrix}
{\bf M}_{ij}(\e)=
\begin{bmatrix}
 {M}_{Ai,Aj}(\e)  &  {M}_{Ai,Bj}(\e)  \\
 {M}_{Bi,Aj}(\e) & {M}_{Bi,Bj}(\e)
\end{bmatrix},
\end{eqnarray}
$\bar {\bf m}_{jj}(\e)=[{\bf I}-{\bf m}_{jj}(\e){\bf V}_{j}]^{-1}{\bf 
m}_{jj}(\e)$ and ${\bf m}_{jj}(\e)=2[\e-\Sigma_{0}(\e)\delta_{0,j}]^{-1} 
{\bf I}$. Here $\Sigma_{0}(\e) \equiv \Sigma_{\rm dot} = 2\sum_\veck 
|\tilde V_{\veck}|^2[(\e-\tilde\e_{\veck})^{-1} + (\e+\tilde\e_{\veck} 
)^{-1}]$ is the dot level broadening (leads), with $\tilde \e_\veck= 
\e_\veck- \e_F$, $V_{S\veck}=V_{D\veck}=\tilde V_{\veck}/\sqrt{2}$ and $\bf 
I$ the $2\times 2$ identity matrix.
Finally,
\begin{eqnarray}\label{coupling}
{\bf V}_{j}=
\frac{1}{2}\begin{pmatrix}
 0 &{i\mu_{j}}\\
-{i\mu_{j}}&0
\end{pmatrix} \, \, \, \, \,  \mbox{and }\, \, \, {{\bf W}_{j}=
\frac{1}{2 }\begin{bmatrix} 
 0 &{iW^{(+)}_{j}}\\
{iW_{j}^{(-)} }&0
\end{bmatrix},}
\end{eqnarray}
with $\mu_{0}=eV_g-2\sum_\veck{|\tilde V_{\veck}|^2[(\e-\tilde 
\e_{\veck})^{-1}-(\e+\tilde\e_{\veck})^{-1}] }$, $W_0^{(\pm)}=\pm t_0$, and 
$\mu_{j}=\mu$ and $W_{j}^{(\pm)}=(\D\pm t$)/2 for all ${j}>0$. The quantity 
$W_{j}^{(\pm)}$  is an effective coupling matrix,  see Fig.~\ref{fig1}(a). 
In the wide band  limit and assuming a constant $\tilde V_{\veck}= 
\sqrt{2}\tilde{V}$ ,  we obtain $\Sigma_{\rm dot}(\e)=-2i\Gamma_L$ and 
$\mu_0=eV_g=-\e_\QD$, with the broadening $\Gamma_L= 2\pi|\tilde{V}|^2 
\rho_L$ and $\rho_L=\rho(\varepsilon_F)$ being the DOS of the leads.
Similarly to (\ref{G11}), we find for the first site ($j=1$) of the chain
\begin{eqnarray}\label{G22}
 {\bf M}_{11}(\e)=\tilde{\bf m}_{11}(\e)+\tilde{\bf m}_{11}(\e){\bf
W}_1^\dagger{\bf M}_{21}(\e),
\end{eqnarray}
with $\tilde{\bf m}_{11}(\e)=\left[{\bf I}-\bar{\bf m}_{11}(\e){\bf
W}_0\bar {\bf m}_{00}(\e){\bf W}_0^\dagger\right]^{-1}\bar{\bf m}_{11} 
(\e)$.
We can then recursively obtain the Majorana matrix at any site.
\section{Numerical results}\label{Results}
Following realistic simulations~\cite{PhysRevB.87.024515,Elsa} and 
experiments,\cite{Science.336.1003} here we assume 
$t=10\meV$, the dot level broadening $\Gamma_L=4.0\times 
10^{-3}t=40\mueV$ and set $\e_F=0$ (we also set $\phi=0$). 
In Fig.~\ref{fig1}(b) we show the LDOS as a function of the
energy $\e$ for a site in the middle and on the edge of 
the chain, $\rho_\text{``bulk"}$ [dashed (red) curve] and 
$\rho_1=\rho_\text{edge}$ [solid (black)  curve], respectively, for $t_0=0$ 
(decoupled chain) and $\D=0.2t=2\meV$. Note that $\rho_{\text{``bulk"}}$ is 
fully gapped, while $\rho_1=\rho_\text{edge}$ exhibits a midgap zero-energy 
peak, corresponding to the end Majorana state of the chain.

Figures \ref{fig1}(c) and \ref{fig1}(d) show the LDOS of the dot  
$\rho_\text{dot}$ and of the first chain site $\rho_1$ as functions of 
$\varepsilon$ for $\e_\QD=-5\Gamma_L$ and three different values of $t_0$. 
For clarity, the curves are offset vertically. For $t_0=0$ [long dashed 
(black) line] we see just the usual single particle peak of width 
$\Gamma_L$ centered at $\e=\e_\QD$. Observe that there is essentially no 
density of states at $\e=0$, since the dot level is far below the Fermi 
level of the leads. As we increase $t_0$ to $2\Gamma_L$ [fine solid (red) 
line], however, we observe the emergence of a sharp peak at $\e=0$ in 
addition to the peak at $\e\approx\e_\QD$. For $t_0=10\Gamma_L$ [dashed 
(blue) line in Fig.~\ref{fig1}(c)], the single-particle  peak in $\rho_{\rm 
dot}$ slightly moves to lower energies while its zero-energy peak increases 
to $0.5$ (in units of $\pi\Gamma_L$). As the this peak appears in 
$\rho_\QD$ for increasing $t_0$'s, the Majorana central peak in 
Fig.~\ref{fig1}(d) decreases. We can still  see a peak in $\rho_1$ for 
$t_0=10\Gamma_L$, dashed  (blue) line in Fig.~\ref{fig1}(d), but much 
weaker than its $t_0=0$ value. We further show $\tilde\rho_\QD= 
\rho_\QD(0)/\rho^{\rm max}_\QD$ and $\tilde \rho_1=\rho_1(0)/\rho^{\rm 
max}_1$, $\rho^{\rm max}_{\rm dot,1}={\rm max}[\rho_{\rm dot,1}(\e=0, 
t_0)]$, vs  $t_0$ in Fig.~\ref{fig1}(e) clearly showing the wire Majorana 
leakage into the dot.

In Fig.~\ref{fig1}(f) we display a color map of the electronic LDOS  
$\rho_\text{dot}$ vs $\e$ and $eV_g$ for the wire in the topological 
phase ($\Delta>0$ and $|\mu|<t$) with $\mu=0$. At $eV_g=0$  we see three 
bright regions that correspond to the three peaks of  $\rho_\text{dot}$ vs. 
$\e$ Ref.~\onlinecite{PhysRevB.84.201308}. In contrast, by fixing $\e=0$ 
and 
varying $eV_g$, we see that the zero-energy peak remains essentially 
unchanged over the range of $eV_g$ shown. More strikingly, this central 
peak is pinned at $\varepsilon=\e_F=0$ for $eV_g>0$  and  $eV_g<0$. The 
pinning for $\e_{\rm dot}$ below $\e_F=0$ is similar to that of the Kondo 
resonance, which, however, is known to occur at $\pi\Gamma_L$,  cf Figs. 
1(h) and 1(i).

\begin{figure}
\centerline{\resizebox{3.6in}{!}{
\includegraphics{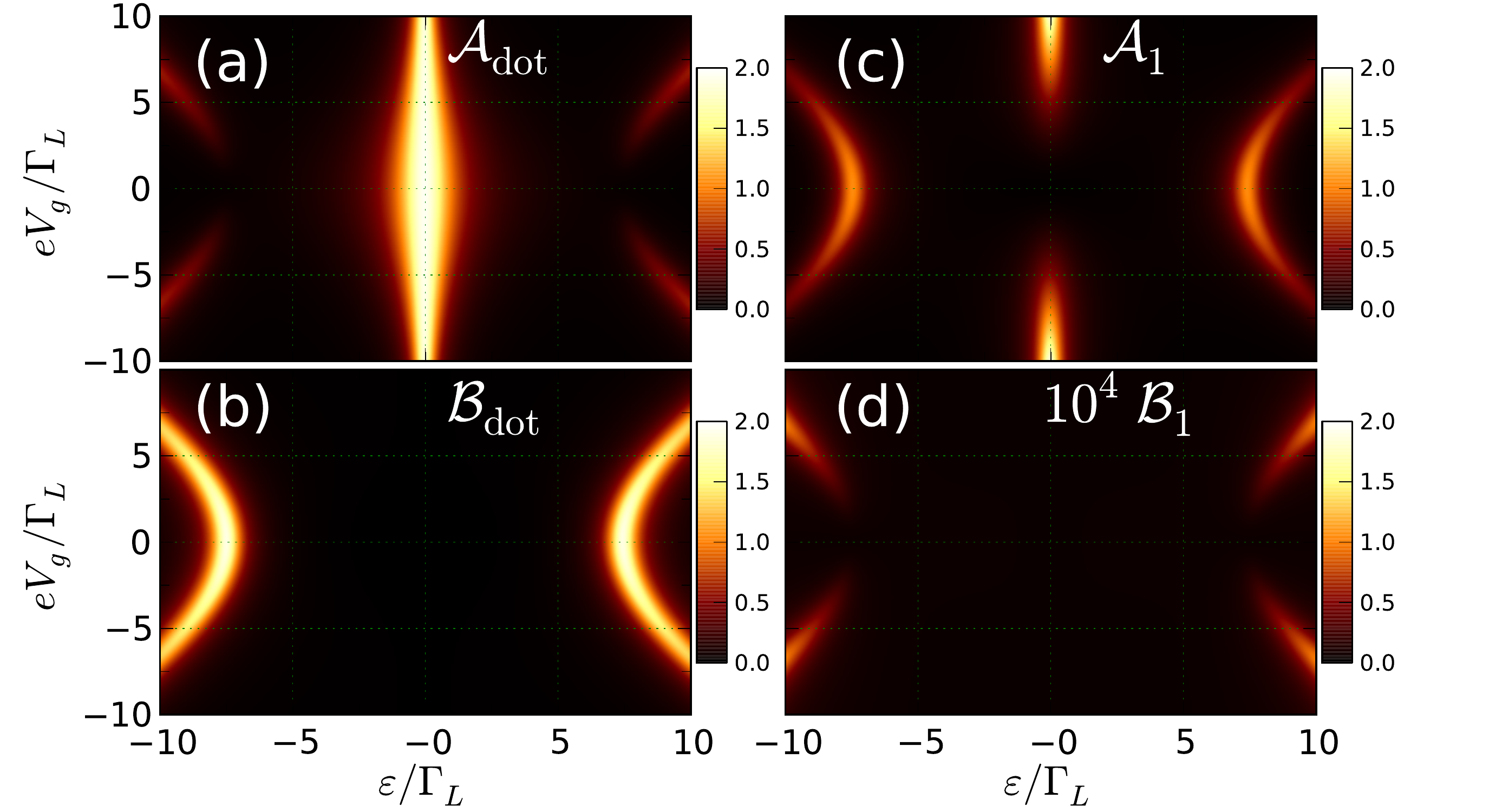}}}
\caption{\label{fig2} (Color online) Color map of the local 
density of states for Majoranas ``A'' (top) and ``B'' (bottom)  at the 
dot (left) and at the first site of the chain (right), as a function of 
$\e$ and $eV_g$ for $t=10\meV$, $\Delta=0.2t$, $\Gamma_L=40\mueV$,  
$t_0=10\Gamma_L$ and $\mu=0$. Panel (d) shows $10^4{\cal B}_1$.  }
\end{figure}

Here again one can measure $G$ vs $V_g$ [Fig.~\ref{fig1}(g)]: for 
the wire in its trivial phase ($|\mu|>t$), e.g.,  $\mu=1.5t$ [circles 
(black)], $G$ exhibits a single peak, whose maximum corresponds to $\e_{\rm 
dot}(V_g)$ crossing the Fermi level. Note that the peak is not at $eV_g=0$ 
but slightly shifted. This arises from the small real part of the self 
energy in the dot Green's function. In the topological phase ($|\mu|<t$), 
e.g.  $\mu=0$ and $\mu=0.75t$ [squares (red) and diamonds (blue), 
respectively], we see an almost constant  $G \simeq e^2/2h$ for $eV_g$ up 
to $\pm10\Gamma_L$. This conductance plateau is similar to that of 
Kondo,\cite{Nature.391.156} except that here $G$ is half of it (per spin) 
and  the plateau occurs even for $\e_\QD>\e_F$.

The Majorana LDOS ${\cal A}_\QD$ and ${\cal B}_\QD$ shown in 
Figs.~\ref{fig2}(a) and \ref{fig2}(b), respectively, as functions of $\e$ 
and $eV_g$ [same parameters as in Fig.~\ref{fig1}(f)], display a 
zero-energy peak  in ${\cal A}_\QD$ and none in ${\cal B}_\QD$. This shows 
that the pinned dot-Majorana peak in Fig.~\ref{fig1}(f) arises from 
Majorana A only. We note that the peaks in ${\cal B}_\QD$ at $\e \approx 
\pm 7\Gamma_L$ (for $eV_g=0$) are affected by the dot-wire Majorana 
coupling as compared to the $\D=t$ case. For couplings to any ordinary 
fermionic wire modes, the dot LDOS would obey ${\cal A}_\QD={\cal B}_\QD$ 
and it would split and broaden.
 
Figures \ref{fig2}(c) and \ref{fig2}(d) show that the Majorana LDOS of the 
first chain site ${\cal A}_1$ and ${\cal B}_1$ have no zero-energy peaks, 
thus indicating that the wire end mode has indeed leaked into the dot. We 
see two peaks in ${\cal A}_1$ at $\e=\pm 7\Gamma_L$ [see 
Fig.~\ref{fig2}(b)] resulting from the coupling $\sim t_0$  between $A_1$  
and $B_{\rm dot}$, see Fig.~\ref{fig1}(a). A careful look at 
Fig.~\ref{fig2}(c) reveals an enhancement of the zero-energy peaks for 
$eV_g\gtrsim 5\Gamma_L$, as a result of the coupling between the dot 
Majorana A and the Majoranas of the chain via a finite $\e_\QD$. The 
strength  of this peak is much smaller than its magnitude without the dot.
\begin{figure}
\centerline{\resizebox{3.4in}{!}{
\includegraphics{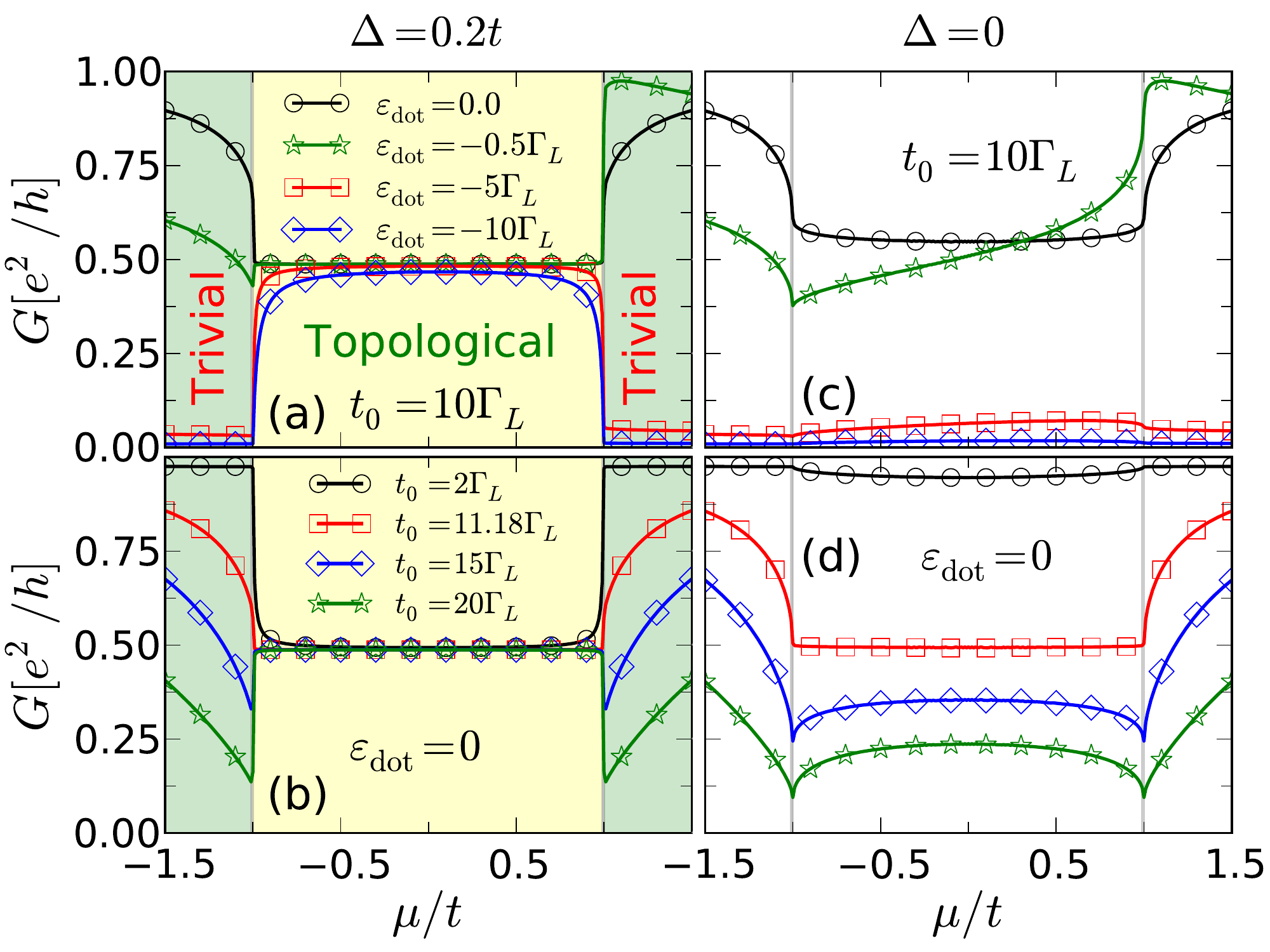}}}
\caption{\label{fig3} (Color online) Conductance $G$ as a 
function of $\mu$ for $t=10\meV$, $\Delta=0.2\meV$ and (a) $t_0=10\Gamma_L$ 
 and different values of $\e_\QD$ and (b) $\e_\QD=0$ and distinct $t_0$'s. 
The lighter (yellow) and darker (green) regions in (a) and (b) highlight 
the topological ($|\mu|<t$) and trivial ($|\mu|>t$) phases  of the  chain, 
respectively. Panels (c) and (d) correspond to (a) and (b), respectively, 
but for $\Delta=0$.} 
\end{figure}

Figure~\ref{fig3}(a) shows the conductance $G$ vs $\mu$ for several 
$\e_\QD$'s [same parameters as in Figs.~\ref{fig1}(f) and \ref{fig1}(g)]. 
For $\e_\QD=0$ [circles (black)] and $|\mu|>t$ (trivial phase), $G$ arises 
from the single-particle dot level at $\varepsilon_F$. The effect of the 
chain is essentially to shift and broaden $\e_\QD$, so that the value 
$e^2/h$ is reached only for $|\mu|\gg t$. As $\mu$ varies across $\pm t$, 
the wire undergoes a trivial-to-topological transition and $G$ suddenly 
decreases to $e^2/2h$ as the leaked dot Majorana appears. For $\e_\QD\neq 
0$ the asymptotic ($|\mu|\gg t$ ) value of the $G$ is no longer $e^2/h$ as 
$\e_\QD$ cannot attain $\varepsilon_F$. The  squares (red) and diamonds 
(blue) in Fig.~\ref{fig3}(a) show a tiny conductance for $\mu>t$. However, 
as $|\mu|$ becomes smaller than $t$ both curves rapidly go to $e^2/2h$.

In Fig.~\ref{fig3}(b) we fix $\e_\QD=0$ and plot the conductance $G$ 
 as a function of $\mu$ for distinct $t_0$'s. As $t_0$ increases, $G$ 
remains pinned at $e^2/2h$ in the topological regime, while it decreases in 
the trivial phase since the dot level shifts due to the chain self energy 
$\sim t_0^2$. Figures \ref{fig3}(c) and \ref{fig3}(d) show $G$ for 
$\Delta=0$ and the same parameters as in Figs.~\ref{fig3}(a) and 
\ref{fig3}(b), respectively. For $|\mu|<t$ the $G$  is very sensitive to 
$\e_\QD$ for a fixed $t_0=10\Gamma_L$  [Fig.~\ref{fig3}(c)], and to $t_0$ 
for $\e_\QD=0$ [Fig.~\ref{fig3}(d)], which contrasts with its practically 
constant value for $\Delta=0.2t$, Figs.~\ref{fig3}(a)  and \ref{fig3}(b). 
This is so because the  wire acts as a third normal lead for $\D=0$ and 
$t_0\neq0$, so the source-drain $G$, e.g., for $\mu=0$, reduces to 
$G=(e^2/h)\Gamma_L/(\Gamma_L+\Gamma_{\rm chain})$, where $\Gamma_{\rm 
chain} =2t_0^2/t$ is the broadening due to the chain.\cite{footnote3} 
Curiously, for $t_0=11.18\Gamma_L$ and $\varepsilon_{\rm dot}=0$ the $G$ 
curves are indistinguishable for $\Delta=0$ and $\D\neq0$, being pinned at 
$e^2/2h$ in the topological {\it and} trivial phases, cf. squares in 
\ref{fig3}(d) and~\ref{fig3}(c). 
{Therefore, the peak value $G=e^2/2h$, first found in Ref. 
\onlinecite{PhysRevB.84.201308} in a similar setup as ours but only for an 
on-resonance dot (i.e., $\varepsilon_{\rm dot}=0=\varepsilon_F$), 
\textit{is not} per se a `smoking-gun' evidence for a Majorana end mode
in conductance measurements, as we find that this peak value can appear 
even in the trivial phase of the wire. One should vary, e.g., 
$\varepsilon_{\rm dot}$ and/or $t_0$ to tell these phases apart as we do in
Fig.~\ref{fig3}.}
Finally, the kinks in 
\ref{fig3}(d) [e.g., diamonds (blue) and stars (green)] result from 
discontinuities in $\Sigma_{\rm chain}$~\cite{footnote3} at $\mu=\pm t$.

\section{Concluding remarks}\label{Remarks} We have used an exact 
recursive 
Green's function approach to calculate the LDOS and the two-terminal 
conductance $G$ through a quantum dot side-coupled to a Kitaev wire. 
Interestingly, we found that the end Majorana mode of the wire leaks into 
the quantum dot thus originating a resonance pinned to the Fermi level of 
the leads $\e_F$. In contrast to the usual Kondo resonance arising only for 
$\e_{\rm dot}$ below $\e_F$, this unique dot-Majorana resonance appears 
pinned to $\e_F$ even when the gate-controlled energy level $\e_{\rm 
dot}(V_g)$ is far above or below $\e_F$, provided that the wire is in 
its topological phase. This leaked Majorana dot mode provides a clear-cut 
way to probe the Majorana mode of the wire via conductance measurements 
through the dot. 

\begin{acknowledgements}
We acknowledge helpful discussions with F. M. Souza.
J.C.E. also acknowledges valuable discussions with D. Rainis.
This work was supported by the Brazilian agencies CNPq,
CAPES, FAPESP, FAPEMIG, and PRP/USP within the Re-
search Support Center Initiative (NAP Q-NANO).
\end{acknowledgements}


\begin{thebibliography}{revtex4-1}
\bibitem{mapping-U} We have also implemented a spin-full  version of our 
system, considering a tight-binding nanowire + Rashba spin-orbit 
interaction + proximity-induced superconductivity + a Zeeman field (a wire 
with these ingredients can be mapped onto the Kitaev 
chain,\cite{PhysRevB.81.125318,NatPhys412}) which corroborates our 
findings. In addition, We have considered a Hubbard $U$ term in the dot 
and 
verified that our results still hold, provided that only a single 
spin-split dot level lies below the Fermi level of the leads (i.e., 
Coulomb 
blockade is irrelevant in this regime). These results will be described 
elsewhere.
\bibitem{footnote3} The general expression for the conductance 
at $T=0$ is $G =\Gamma_L \left(\Gamma_L - \text{Im }\Sigma_{\rm 
chain}\right)/
[\left(\text{Re }\Sigma_{\rm chain}\right)^2 + \left(\Gamma_L - \text{Im 
}\Sigma_{\rm chain}\right)^2]$, with $\Sigma_{\rm chain}=2t_0^2\left(1 - 
\sqrt{1-g_0^2t^2}\right)/g_0t^2$ and $g_0=(-\mu+i\eta)^{-1}$.
\end{thebibliography}
\end{document}